\documentclass[submission, Proceedings]{SciPost}

\hypersetup{
    colorlinks,
    linkcolor={red!50!black},
    citecolor={blue!50!black},
    urlcolor={blue!80!black}
}

\def\Pomeron{I\!\!P}

\usepackage[bitstream-charter]{mathdesign}
\DeclareSymbolFont{usualmathcal}{OMS}{cmsy}{m}{n}
\DeclareSymbolFontAlphabet{\mathcal}{usualmathcal}

\begin{document}

\begin{center}{\Large \textbf{
Inclusive and diffractive dijet photoproduction in ultraperipheral heavy ion collisions at the LHC
\\
}}\end{center}

\begin{center}
V. Guzey\textsuperscript{1$\star$}
and
M. Klasen\textsuperscript{2}
\end{center}

\begin{center}
{\bf 1} National Research Center ``Kurchatov Institute'', Petersburg Nuclear Physics Institute (PNPI), Gatchina, 188300, Russia
\\
{\bf 2} Institut f\"ur Theoretische Physik, Westf\"alische Wilhelms-Universit\"at M\"unster, Wilhelm-Klemm-Stra{\ss}e 9, 48149 M\"unster, Germany
\\
* guzey\_va@pnpi.nrcki.ru
\end{center}

\begin{center}
\today
\end{center}


\definecolor{palegray}{gray}{0.95}
\begin{center}
\colorbox{palegray}{
  \begin{tabular}{rr}
  \begin{minipage}{0.1\textwidth}
    \includegraphics[width=22mm]{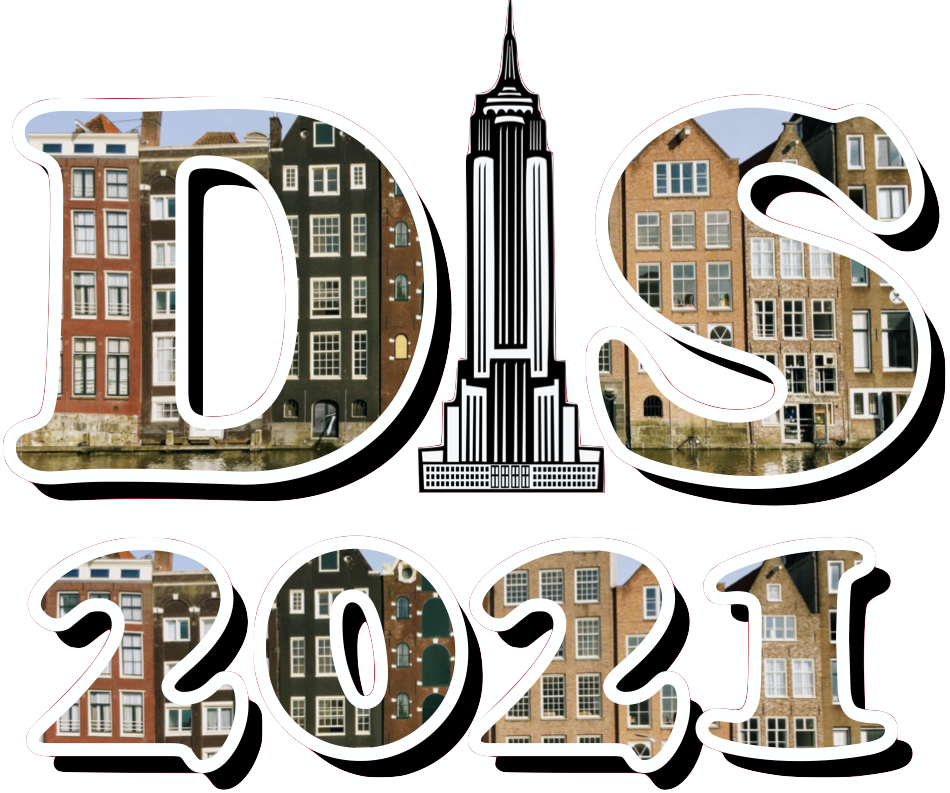}
  \end{minipage}
  &
  \begin{minipage}{0.75\textwidth}
    \begin{center}
    {\it Proceedings for the XXVIII International Workshop\\ on Deep-Inelastic Scattering and
Related Subjects,}\\
    {\it Stony Brook University, New York, USA, 12-16 April 2021} \\
    \doi{10.21468/SciPostPhysProc.?}\\
    \end{center}
  \end{minipage}
\end{tabular}
}
\end{center}

\section*{Abstract}
{\bf
We calculate the cross section of inclusive dijet photoproduction in ultraperipheral collisions (UPCs) of heavy ions at the CERN Large Hadron Collider using next-to-leading order perturbative QCD and demonstrate that it provides a good description of the ATLAS data. We study the role of this data in constraining nuclear parton distribution functions (nPDFs) using the Bayesian reweighting technique and find that it can reduce current uncertainties of nPDFs at small $x$ by a factor of 2. We also make predictions for diffractive dijet photoproduction in UPCs and examine its potential to shed light on the disputed mechanism of QCD factorization breaking in diffraction.
}


\section{Introduction}
\label{sec:intro}

In ultraperipheral collisions (UPCs), ions in colliding beams intersect at large impact parameters (large transverse separations), $b \gg R_A+R_B$, where $R_A$ and $R_B$ denote the radii of the involved ions. In this case,
the short-range strong hadron interactions are suppressed and the reaction proceeds through emission of quasi-real photons, which is usually treated in the 
Weizs\"acker--Williams equivalent photon approximation \cite{Budnev:1974de}.
Since the intensity of the flux of these photons scales as $Z^2$, where $Z$ is the electric charge of the photon-emitting ion, and the maximal photon energy in the laboratory frame is proportional to $\gamma_L$, where $\gamma_L$ is the ion Lorentz factor, UPCs in the kinematics of the Large Hadron Collider (LHC) allow one to study photon--photon, photon--proton, and 
photon--nucleus interactions at unprecedentedly high energies \cite{Bertulani:2005ru,Baltz:2007kq,Contreras:2015dqa}. 
In particular, studies of UPCs at the LHC enable one to can address open questions of the proton and nucleus structure in QCD and also search for signs of new physics.

The focus of UPC measurements at the LHC has been exclusive photoproduction of charmonia 
($J/\psi$, $\psi^{\prime}$) and light  $\rho$ vector mesons. The former provided new constraints on the
gluon density at small momentum fractions $x$ down to $x_p \sim 6 \times 10^{-6}$ in the proton case and down to $x_A \sim 6 \times 10^{-4}$ in the case of lead (Pb) nuclei \cite{Adam:2020mxg}.
In addition, 
the presently poorly constrained nuclear parton distribution functions (PDFs) \cite{Kovarik:2015cma,Eskola:2016oht} and photon PDFs \cite{Gluck:1991jc} can also be accessed using complimentary processes such as, e.g., inclusive dijet photoproduction in Pb-Pb UPCs \cite{Guzey:2018dlm,Guzey:2019kik}.
Besides, requiring that the nuclear target remains intact, one can study diffractive dijet photoproduction in Pb-Pb UPCs, which measures novel nuclear diffractive PDFs and may help to determine the mechanism of QCD factorization breaking in diffractive scattering \cite{Guzey:2016tek}.

\section{Inclusive dijet photoproduction in Pb-Pb UPCs at the LHC}
\label{sec:inc_dijet_upc}

\begin{figure}[t]
\centering
\includegraphics[width=0.7\textwidth]{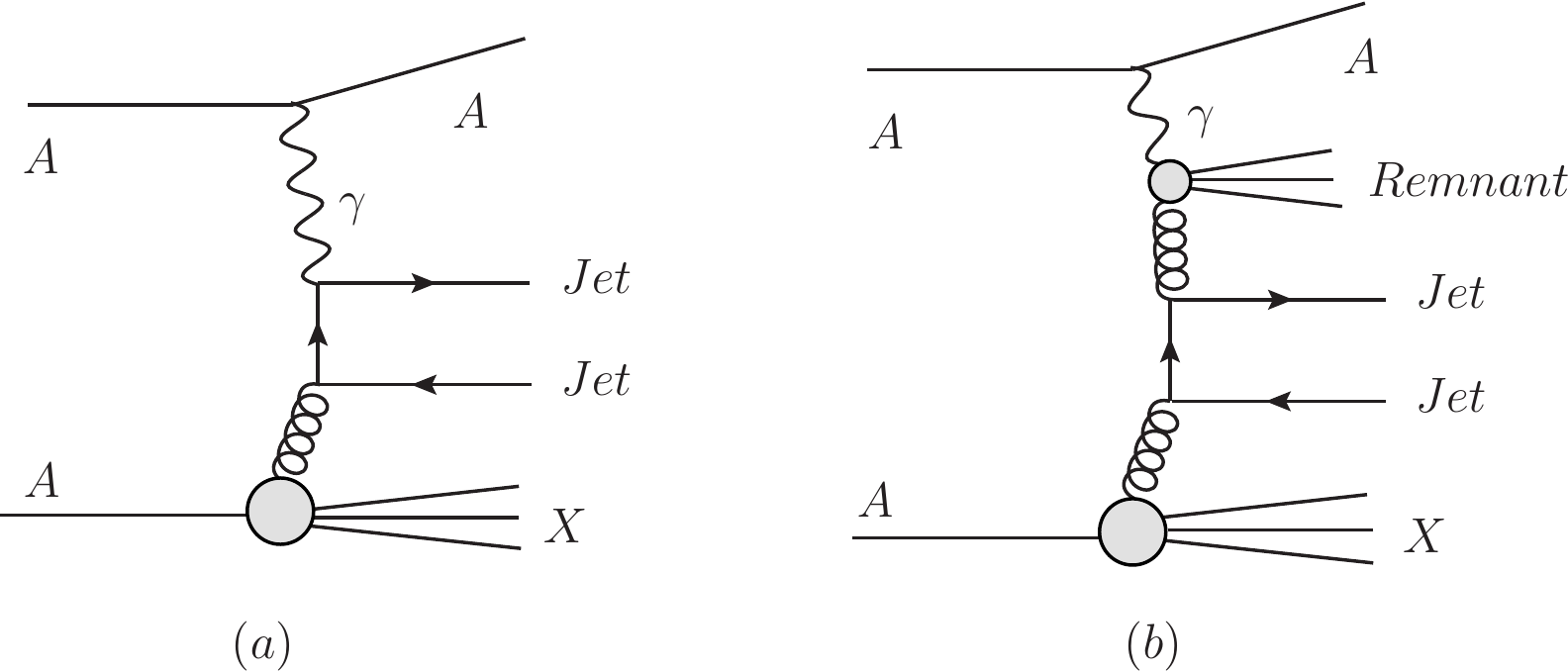}
\caption{Typical leading-order Feynman graphs for dijet photoproduction in UPCs. Graphs (a) and (b) correspond to the direct and resolved photon contributions, respectively.}
\label{fig:photo_jets_upc2}
\end{figure}

Inclusive dijet photoproduction on the proton was extensively studied at the Hadron--Electron Ring Accelerator (HERA) \cite{Newman:2013ada} and various measured distributions were successfully described in the framework of collinear factorization and next-to-leading order (NLO) perturbative QCD \cite{Klasen:2002xb,Klasen:2011ax}. Generalizing this to the case of heavy ion UPCs, the cross section of dijet photoproduction in Pb-Pb UPCs (see Fig. \ref{fig:photo_jets_upc2}) can be written in the following form \cite{Guzey:2018dlm}
\begin{eqnarray}
&&d\sigma(AA \to A+{\rm 2 jets}+X) = \nonumber\\
&& \sum_{a,b} \int dy \int dx_{\gamma} \int dx_A f_{\gamma/A}(y)f_{a/\gamma}(x_{\gamma},\mu^2)f_{b/A}(x_A,\mu^2) d\hat{\sigma}(ab \to {\rm jets})\,,
\label{eq:cs}
\end{eqnarray}
where $a,b$ are parton flavors; $f_{\gamma/A}(y)$ is the flux of equivalent photons;
$f_{a/\gamma}(x_{\gamma},\mu^2)$  and $f_{b/A}(x_A,\mu^2)$ are PDFs of the photon \cite{Gluck:1991jc}
(in the resolved photon case) and the target nucleus \cite{Kovarik:2015cma,Eskola:2016oht}, respectively;
$y$, $x_{\gamma}$, and $x_A$ are longitudinal momentum fractions carried by the photon, partons in the photon, and partons in the nucleus, respectively; finally, $d\hat{\sigma}(ab \to {\rm jets})$ is the elementary cross section for production
of jets in hard scattering of partons $a$ and $b$.
The sum over $a$ involves quarks and gluons for the resolved photon contribution 
and the photon for the direct photon contribution dominating at $x_{\gamma} \approx 1$.

In our analysis, for the photon flux $f_{\gamma/A}(y)$ we used the standard expression 
 \begin{equation}
 f_{\gamma/A}(y)=\frac{2 \alpha_{\rm e.m.}Z^2}{\pi}\frac{1}{y} \left[\zeta K_0(\zeta)K_1(\zeta)-\frac{\zeta^2}{2}(K_1^2(\zeta)-K_0^2(\zeta)) \right] \,,
 \label{eq:flux}
 \end{equation}
where $\alpha_{\rm e.m.}$ is the fine-structure constant; $Z$ is the electric charge;
$K_{0,1}$ are modified Bessel functions of the second kind;
$\zeta=y m_p b_{\rm min}$ with $m_p$ being the proton mass and $b_{\rm min}$ the minimal distance between the two nuclei.
For Pb-Pb UPCs, Eq.~(\ref{eq:flux}) with $b_{\rm min}=14.2$ fm reproduces very well the photon flux calculated taking into account the nuclear form factor and the suppression of strong interactions for impact parameters $b < b_{\rm min}$.

\begin{figure}[t]
\centering
\includegraphics[width=0.45\textwidth]{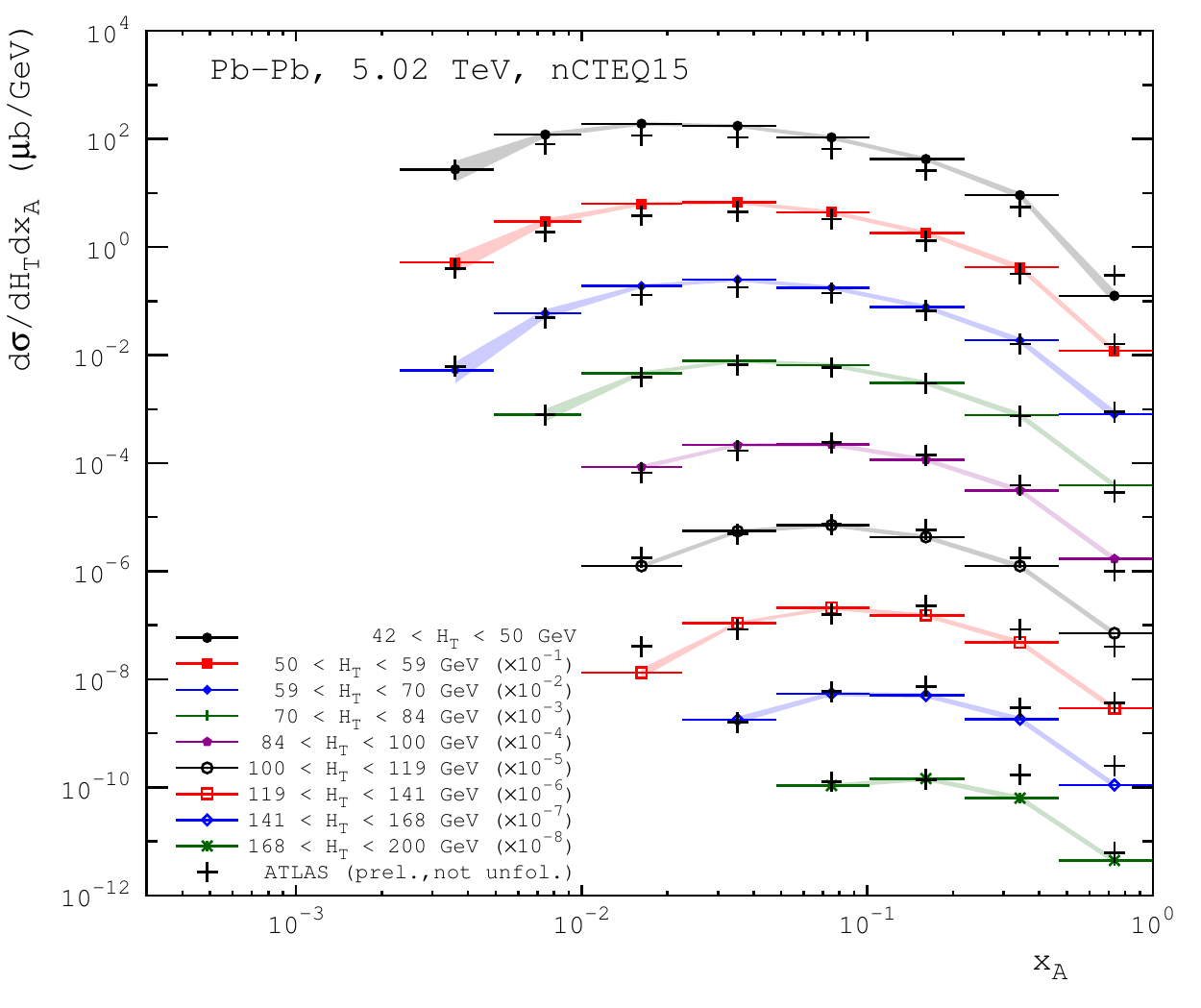}
\includegraphics[width=0.45\textwidth]{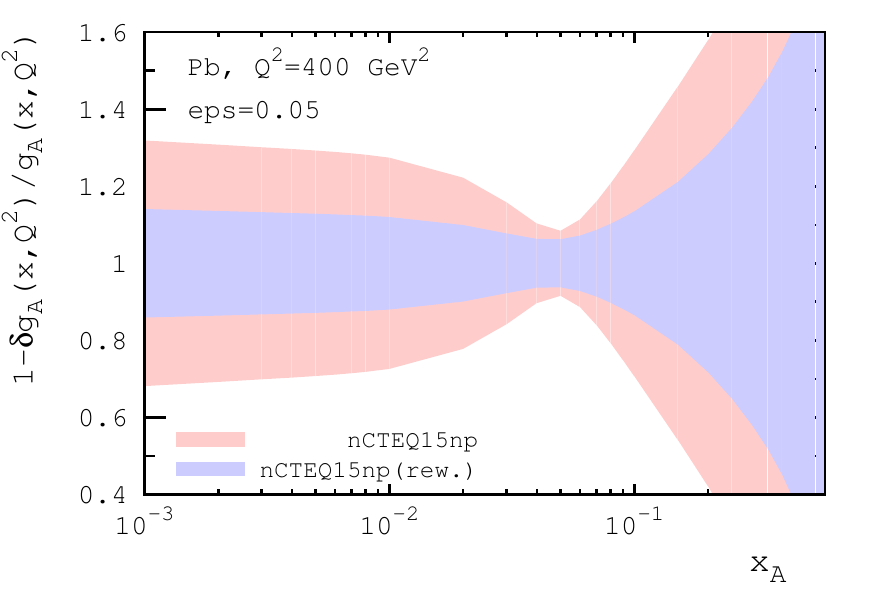}
\caption{(Left) NLO QCD results for the cross section of dijet photoproduction in Pb-Pb UPCs at $\sqrt{s_{NN}}=5.02$ TeV as a function of $x_A$ for different bins of $H_T$ \cite{Guzey:2018dlm}.
The crosses are the ATLAS data points extracted from \cite{Atlas}.
(Right) The relative uncertainty bands of the nCTEQ15np gluon distribution in Pb at $Q^2=400$ GeV$^2$ as a function of $x_A$ before (outer shaded band) and after (inner shaded band) the re-weighting assuming a 5\% error on the pseudo-data \cite{Guzey:2019kik}. 
}
\label{fig:xA_unc_final_full_new}
\end{figure}

Figure \ref{fig:xA_unc_final_full_new} (left) shows the results of the NLO pQCD formalism outlined above
for the cross section of dijet photoproduction in Pb-Pb UPCs at $\sqrt{s_{NN}}=5.02$ TeV as a function of
$x_A$ in different bins of $H_T=p_{T,1}+p_{T,2}$ and implementing the cuts of the ATLAS 
measurement \cite{Atlas} ($p_{T,1} > 20$ GeV for the leading jet and $p_{T,i\neq1}> 15$ GeV for sub-leading ones;
the rapidity interval of $|\eta_i| < 4.4$). 
The shown results are obtained using the anti-k$_T$ algorithm with the jet radius $R=0.4$, nCTEQ5 nuclear PDFs \cite{Kovarik:2015cma}, 
and $\mu=2E_{T,1}$ to have numerical stability against higher-order corrections.
They are compared to the preliminary ATLAS data, which have not been corrected for 
detector response. One can see from the figure that NLO pQCD reproduces well the shape and normalization of the ATLAS data. A similarly good description of other kinematic distributions, e.g., the distributions in $H_T$
and $z_{\gamma}=y x_{\gamma}$ 
has also been achieved \cite{Guzey:2018dlm}.

The potential of inclusive dijet photoproduction in Pb-Pb UPCs to provide complementary constraints on nPDFs
 can be quantified using the statistical method of Bayesian reweighting \cite{Guzey:2019kik}. 
An example of such an analysis is presented in Fig. \ref{fig:xA_unc_final_full_new} (right), which shows the relative uncertainty bands of the nCTEQ15np gluon distribution in Pb at $Q^2=400$ GeV$^2$ as a function of $x_A$ before (outer shaded band) and after (inner shaded band) the reweighting. Assuming a 5\% error on the data 
($\epsilon=0.05$), one can see from the figure that the sufficiently precise measurement of inclusive dijet photoproduction in Pb-Pb UPCs in the LHC kinematics can reduce current uncertainties of nPDFs at small $x$ by a factor of 2.

\section{Diffractive dijet photoproduction in Pb-Pb UPCs at the LHC}
\label{sec:diff_dijet_upc}

Requiring that the target nucleus in Fig. \ref{fig:photo_jets_upc2} remains intact, one can study diffractive 
dijet photoproduction in heavy ion UPCs. Since in this case both ions can serve as a source of photons and as a target, the UPC cross section is given by a sum of right-moving and left-moving photon sources
\begin{equation}
d\sigma(AA \to A+{\rm 2 jets}+X+A)=d\sigma(AA \to A+{\rm 2 jets}+X+A)^{(+)}+d\sigma(AA \to A+{\rm 2 jets}+X+A)^{(-)} \,,
\label{eq:cs_diff_sum}
\end{equation}
where the two terms are related by inversion of the signs of the jet rapidities.
Applying the framework of collinear factorization and NLO pQCD, one obtains \cite{Guzey:2016tek}
\begin{eqnarray}
&&d\sigma(AA \to A+{\rm 2 jets}+X+A)^{(+)} =  \sum_{a,b}\int dt \int dx_{\Pomeron} \int dz_{\Pomeron} \nonumber\\
&&\times \int dy \int dx_{\gamma} f_{\gamma/A}(y)f_{a/\gamma}(x_{\gamma},\mu^2)f_{b/A}^{D(4)}(x_{\Pomeron},z_{\Pomeron},t,\mu^2) d\hat{\sigma}(ab \to {\rm jets})\,,
\label{eq:cs_diff}
\end{eqnarray}
where $f_{b/A}^{D(4)}$ are nuclear diffractive PDF, which represent the conditional probability to find in a nucleus parton $b$ with the momentum fraction $z_{\Pomeron}$ with respect to the diffractive exchange (pomeron), which itself carries the momentum fraction $x_{\Pomeron}$ provided the nucleus remained intact 
and experienced the momentum transfer $t$. 

As usual nPDFs, nuclear diffractive PDFs are subject to nuclear modifications. In particular, the model of leading twist nuclear shadowing \cite{Frankfurt:2011cs} predicts their strong suppression (shadowing) with respect to the impulse approximation (IA) estimate,
\begin{equation}
f_{b/A}^{D(4)}(x_{\Pomeron},z_{\Pomeron},t,\mu^2)=R_b A^2 F_A^2(t) f_{b/p}^{D(4)}(x_{\Pomeron},z_{\Pomeron},t=0,\mu^2)
\end{equation}
where $F_A(t)$ is the nuclear form factor;
the suppression factor of $R_b \approx 0.15$ weakly depends on the parton flavor, the momentum fractions 
$x_{\Pomeron}$ and $z_{\Pomeron}$, and the scale $\mu$. 

\begin{figure}[t]
\centering
\includegraphics[width=0.63\textwidth]{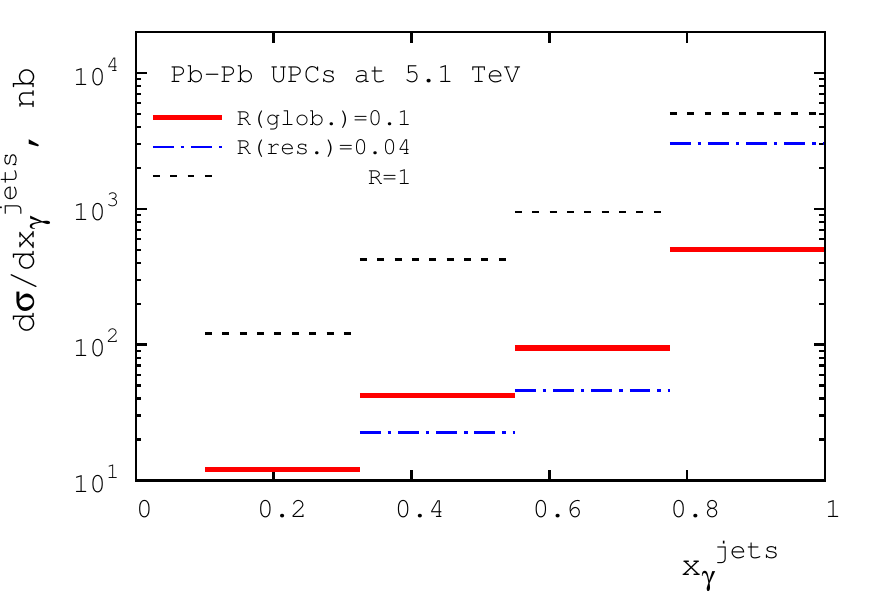}
\caption{NLO pQCD predictions for the cross section of diffractive dijet photoproduction in Pb-Pb UPCs at 5.1 TeV as a function of
 $x_{\gamma}$ for different scenarios of factorization breaking \cite{Guzey:2016tek}. }
\label{fig:AA5_fb_dis}
\end{figure}

Analyses of diffractive dijet photoproduction in electron--proton scattering at HERA have shown that QCD
factorization in diffractive dijet photopropduction is broken, i.e., NLO calculations overestimate the data by approximately a factor of 2. The pattern of this factorization breaking is unknown and is characterized either by a global suppression factor of $R(\rm glob.)=0.5$ or by the resolved-only suppression $R(\rm res.)=0.34$. 
Figure \ref{fig:AA5_fb_dis} shows NLO pQCD predictions \cite{Guzey:2016tek} for the cross section of diffractive dijet photoproduction in Pb-Pb UPCs at 
5.1 TeV as a function of the momentum fraction $x_{\gamma}$, which correspond to different assumptions about the pattern of factorization breaking: no-factorization breaking (dashed), the global suppression with $R(\rm glob.)=0.1$ (dot-dashed), and the resolved-only suppression with $R(\rm res.)=0.04$ (solid). As one can see from the figure, the $x_{\gamma}$ distribution is sensitive to different scenarios of factorization breaking.

Note that the diffractive contribution to inclusive dijet photoproduction in Pb-Pb UPCs in the ATLAS kinematics does not exceed 5\% and, hence, does not affect the analysis of nPDFs at small $x_A$ \cite{Guzey:2020ehb}.

\section{Conclusion}

We calculated the cross sections of inclusive and diffractive dijet photoproduction in UPCs of heavy ions at the LHC using NLO perturbative QCD. We demonstrated that our approach provides a good description of the ATLAS data, which exhibits 10-20\% nuclear modifications. We studied the role of this data in constraining nuclear PDFs using the Bayesian reweighting technique and found that it can reduce current uncertainties of nPDFs at small $x$ by a factor of 2. We also quantified the potential of diffractive dijet photoproduction in UPCs to shed light on the disputed mechanism of QCD factorization breaking in diffraction.

\section*{Acknowledgements}

VG's research is supported in part by RFBR, research project 17-52-12070.  The authors gratefully acknowledge financial support of DFG through the grant KL 1266/9-1 within the joint German--Russian project ``New constraints on nuclear parton distribution functions at small $x$ from dijet production
in $\gamma A$ collisions at the LHC''.







\nolinenumbers

\end{document}